\documentclass[aps,pre,preprint,superscriptaddress]{revtex4-1}

\usepackage{bm}
\usepackage{amsmath}
\usepackage[dvipdfmx]{graphicx}
\usepackage{setspace}
\usepackage{color}
\usepackage{lineno}
\usepackage{array}
\usepackage{txfonts}

\definecolor{forestgreen}{rgb}{0.33,0.61,0.34}

\newcommand{\del}[1]{\textcolor{red}{}}

\begin{document}

\title{Model retraining and information sharing in a supply chain with long-term fluctuating demands}
\author{Takahiro Ezaki*}
\affiliation{Research Center for Advanced Science and Technology, The University of Tokyo, 4-6-1 Komaba, Meguro-ku, Tokyo 153-8904, Japan\\\ *email: tkezaki@g.ecc.u-tokyo.ac.jp}

\author{Naoto Imura}
\affiliation{Research Center for Advanced Science and Technology, The University of Tokyo, 4-6-1 Komaba, Meguro-ku, Tokyo 153-8904, Japan\\ *email: tkezaki@g.ecc.u-tokyo.ac.jp}

\author{Katsuhiro Nishinari}
\affiliation{Research Center for Advanced Science and Technology, The University of Tokyo, 4-6-1 Komaba, Meguro-ku, Tokyo 153-8904, Japan\\ *email: tkezaki@g.ecc.u-tokyo.ac.jp}

\begin{abstract}
Demand forecasting based on empirical data is a viable approach for optimizing a supply chain. However, in this approach, a model constructed from past data occasionally becomes outdated due to  long-term changes in the environment, in which case the model should be updated (i.e., retrained) using the latest data. In this study, we examine the effects of updating models in a supply chain using a minimal setting. We demonstrate that when each party in the supply chain has its own forecasting model, uncoordinated model retraining causes the bullwhip effect even if a very simple replenishment policy is applied. Our results also indicate that sharing the forecasting model among the parties involved significantly reduces the bullwhip effect. 
\end{abstract}

\maketitle

\section{Introduction}
Optimal operation of a supply chain (SC) remains a challenging task. 
A major obstacle to efficient SC management is the intrinsic uncertainty of future demand.
The parties involved in an SC attempt to control the stock level and avoid overproduction and overstocking based on a past 
demand pattern \cite{Chambers1971}. However, when the parties perform this in an uncoordinated manner, 
 fluctuation in demand is amplified and transferred to upstream parties, which is called the bullwhip effect \cite{Lee1997a,Lee1997b,Disney2007}. 
Factors causing the bullwhip effect can be classified into five groups: demand signal processing, lead time, order batching, price fluctuations, and rationing and shortage gaming \cite{Lee1997a,Lee1997b}. 
These inefficient practices amplify the demand signal and transfer it to  upstream parties.

In addition to empirical studies \cite{Cachon1997,Holweg2005,Disney2007}, 
a number of theoretical studies have demonstrated that replenishment policies based on incomplete information (i.e., demand signal processing) are the cause of the bullwhip effect.
Traditional approaches to study inventory control use exponential smoothing \cite{Lee1997a,Lee1997b,Chen2000,Lee2000,Thonemann2002,Zhang2004,Ali2012} and moving average \cite{Chen2000, Zhang2004} techniques. Although these control methods are effective for certain types of demand signals when the strength of the feedback is properly set, feedback that is too strong often destabilizes the SC. Another control approach is a model-based one, in which replenishment is performed based on a model that directly describes the temporal structure of the demands. For example, autoregressive models, including the autoregressive moving average (ARMA) \cite{Disney2006} and autoregressive integrated moving average (ARIMA) \cite{Li2005,Dhahri2007,Ali2012} models, have been used in various studies. These studies demonstrated that information sharing among firms prevents the firms from overreacting to signals and is the key to stabilizing the SC \cite{Raghunathan1999}.

In addition, demand forecasting using machine learning modeling, which aims at high prediction accuracy, has recently attracted attention   \cite{Carbonneau2008,Bajari2015,Makridakis2018,Schelter2018}.
This approach has been facilitated by the increased accessibility of high computational power, state-of-the-art modeling techniques, and big data \cite{Barbosa2018,Nguyen2018,Hofmann2018}. 
Although this approach has not yet achieved satisfactory performance \cite{Makridakis2018}, the use of relatively complex models (e.g., support vector machines  and neural networks \cite{Carbonneau2008, Bajari2015, Makridakis2018, Schelter2018}), which can accommodate more variables, appears promising because the demand may be influenced by many (measurable or unmeasurable) factors \cite{Flynn2016}, such as price \cite{Chen2010}, promotions \cite{Ali2009,Abolghasemi2020}, and calendar events \cite{Huber2020}. Information on these factors is likely to increase the accuracy of the prediction if properly included in the model.

As a feature of mathematical modeling, training (or parameter estimation) based on past data requires higher cost than performing prediction (or inference) based on a model \cite{Januschowski2020}. Thus, a constructed forecasting model is not updated as often as forecasting. 
It should be noted that there is a common implementation in which model parameters are updated (retrained) for each forecast if the cost of doing so is small \cite{Januschowski2020}.
Because the demand pattern may change over a long-term period and is often difficult to model, the forecasting model should be updated at the appropriate time. 
However, the appropriate time for retraining is often unknown for each firm, which may destabilize the SC.

In this study, we examine the scenario in which model retraining is performed in an uncoordinated manner in each firm comprising an SC under long-term variable demand.
We demonstrate that this can cause the bullwhip effect, which can be obviously prevented by sharing the model of the retailer (i.e., the firm at the customer end of the SC) to upstream firms. Information sharing at any level is often difficult to implement in practice due to various reasons, e.g., cost, compatibility of information systems, and confidentiality issues \cite{Ali2017}. 
Thus, evaluating the benefits of sharing the model and cost of not doing so contributes to efficient supply chain management.
We use a simple forecasting model to illustrate this phenomenon, 
excluding the possibility of other factors (e.g., lead time, seasonality, and model structures) confounding the cause of the bullwhip effect.

The remainder of this paper is organized as follows. In Sec. \ref{sec:model}, we define the model and introduce three retraining policies and one baseline (non-retraining) policy. In Sec. \ref{sec:results}, we report the inventory level and the amount of sales loss yielded by each policy. We demonstrate that uncoordinated model retraining causes the bullwhip effect, and that sharing the forecasting model improves the performance of the SC. In Sec. \ref{sec:conclusion}, we summarize the results and discuss their implications.

\section{Model}\label{sec:model}
\subsection{Order and shipment}
\begin{figure*}[t]
\centering\includegraphics[width=160mm]{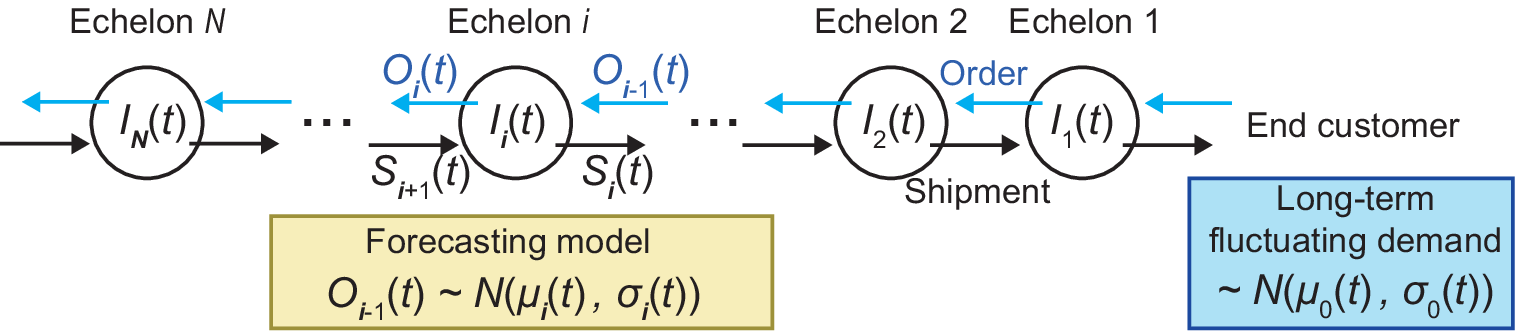}
\caption{Overview of system. Each echelon places an order to the next upstream echelon according to the order-up-to-point policy. 
The target inventory level is determined by the forecasting model in each echelon, which is temporally updated. 
The demand at the end customer is stochastically generated by a probability distribution, $\mathcal{N}(\mu_0(t),\sigma_0(t))$.
}
\label{fig:schematic}
\end{figure*}

We consider a simple SC consisting of $N$ echelons adopting an order-up-to-point policy \cite{Chen2000a, Lee2000, Gilbert2005} (Fig. \ref{fig:schematic}). Each echelon may represent a retailer, wholesaler, or manufacturers of certain consumer goods, for example. When goods are sold at the consumer end of the SC (i.e., retailer), its inventory is reduced, and orders are placed with a wholesaler to replenish the products. The wholesaler places orders with the next wholesaler or manufacture to meet the demand. As the result of these orders, products or materials are sent in the opposite direction.

The target inventory level in the policy is determined by the forecasting model of each echelon, which is described in Sec \ref{subsec:policy}.
Briefly, each echelon first receives an order from a downstream echelon (or the end customer) and performs shipments to it, followed by an immediate replenishment process by sending an order to the next upstream echelon, in a single discrete time step. A single discrete time step represents an interval between successive orders, which may represent a day, week, and so forth, in a real situation. For simplicity, here, we assume that each firm in the SC places an order in a synchronized manner in each time step. Note that we ignore the lead time in this model to exclude its effects on the bullwhip effect in our model.

Technically, we update the system from $t = T$ to $t = T+1$ as follows.  First, at the most downstream customer, demand is generated by a Gaussian distribution, $\mathcal{N}(\mu_{0}(T), \sigma_{0}(T))$ (with temporally variable parameters; see Sec. \ref{sec:demand_fluctuation} for details). Note that we have carried out the entire simulations with log-normal distributions as well (Supporting Information) to confirm the conclusions were not specific to the Gaussian distribution. 
Then, we sequentially update each echelon from the downstream.  
An update of an echelon is twofold, involving a shipment to the next downstream echelon and subsequent replenishment by placing an order to the next upstream echelon. Echelon $i$ receives an order from the next downstream echelon $i-1$ (or the end customer for $i=1$), which is denoted by $O_{i-1}(T)$. If the ordered amount is smaller than the inventory level of echelon $i$, $I_{i}(t)$ (i.e., $O_{i-1}(T) < I_{i}(T)$), echelon $i$ depletes the ordered amount (i.e., $I_i(T+0.5) = I_i(T) - O_{i-1}(T)$); otherwise, it depletes the entire current inventory (i.e., $I_i(T+0.5) = 0$). The downstream echelon $i-1$ receives this shipment from echelon $i$, $S_{i}(T) = \max{\{O_{i-1}(T), I_{i}(T)\}}$, resulting in $I_{i-1}(T+1) = I_{i-1}(T+0.5) + S_i(T)$. For $i=N$, the ordered amount is replenished (i.e., $I_{N}(T+1) =I_{N}(T+0.5) +O_N(T)$). Note that we do not consider the backlog of orders; thus, the amount of demand that exceeds the inventory level is lost, which is recorded as the lost sales opportunity.

\subsection{Order policy and models for demands}\label{subsec:policy}
Each echelon $i$ has a forecasting model for demand as a probability distribution, based on which it determines the safety stock and target inventory levels. Here we use the Gaussian distribution,  $O_{i-1}(t) \sim \mathcal{N}(\mu_i(t), \sigma_{i}(t))$, as the forecasting model.  
This model is one of the simplest forecasting models and mimics the functions of other types of more complex models that generate a prediction with an estimated error. We use this model to exclude the possibility that a specific temporal structure of a forecasting model, not the retraining of the models on which we focus here, causes the bullwhip effect. Essentially the same models for the demand have been widely used in the literature to theoretically investigate the dynamics of SCs \cite{Cachon2000, Raju2000, Cachon2001, Kim2006}.

Using the two parameters in the model (i.e., $\mu_i$ and $\sigma_i$ denoting the mean and standard deviation, respectively), the target inventory level is set to 
\begin{equation}
I_{i,\rm{target}}(t) = \mu_i(t) + c \sigma_i(t),\label{eq:target}
\end{equation} 
where $c$ is the safety factor defining the amount of safety stock (i.e., order-up-to-point policy \cite{Chen2000a, Lee2000, Gilbert2005}).
The ordered amount is simply the difference between the target and  current (at $t = T+0.5$) inventory levels (i.e., $O_i(T) = I_{i,\rm{target}}(T) - I_{i}(T+0.5)$). If the current inventory level is higher than the target inventory level (which may occur when the forecasting model is retrained), we set $O_i(T) = 0$. Note that this order policy alone does not amplify the demand signal, and thus does not cause the bullwhip effect \cite{Ren2017}.

\subsection{Fluctuations in demand} \label{sec:demand_fluctuation}
We also fluctuate the demand distribution at the end customer, $\mathcal{N}(\mu_0,\sigma_0)$. 
As an example, we set the following simple stochastic process with moderate variability.
We fixed $\sigma_0 = 0.1$. When updated, the mean of the distribution, $\mu_0$, was redrawn from a uniform distribution between $1/2$ and $2$, i.e., $\mu_{0, {\rm new}} \sim U(1/2, 2)$. 

We considered two types for intervals of updating the distribution, $L_{\rm{int}}$. The first is a constant update interval, with which the demand distribution is updated every $L_{\rm{int}}(=Const.)$ time steps. In this study, we examined $L_{int} = 50$ and $100$. The second type of interval is a random update interval, which is redrawn from the uniform distribution, $U(L_{\rm{min}},L_{\rm{max}})$, every time the demand distribution is updated. Here, we set $L_{\rm{min}} = 50$ and $L_{\rm{max}}=100$. These two update intervals were used to show that our results were not influenced by either discrete nor stochastic properties of the demand.

\subsection{Retraining schemes for demand forecasting at echelons}
As the demand pattern varies, the forecasting models of the echelons are also updated (i.e., retrained). Each echelon $i$ refers to the past $L_{\rm{train}}$ orders it received, $\{O_{i-1}(T-1),\ldots,O_{i-1}(T-L_{\rm{train}})\}$. When the model is updated, $\mu_i$ and $\sigma_i$ were replaced by the mean and corrected standard deviation computed from the sample, respectively. The optimal value of $L_{\rm{train}}$ depends on the environment and is generally unknown.

We consider the following four types of schemes defining when and how to perform retraining at each echelon.

\subsubsection{Regular update}
We simultaneously update the models of all echelons every $L_{\rm{train}}$ steps, i.e., at $t = nL_{\rm{train}}$ $(n=1,2,...)$. 

\subsubsection{Independent update}
At every time step, each echelon $i$ verifies whether the sample average computed from the most recent past data, $\{O_{i-1}(T-1),\ldots,O_{i-1}(T-L_{\rm{train}})\}$, falls within the interval $[\mu_i - 1.96\sigma_i / \sqrt{L_{\rm{train}}}, \mu_i + 1.96\sigma_i/ \sqrt{L_{\rm{train}}}]$, and if not, it updates the forecasting model. If the current model correctly predicts the demand, 95\% of the sample average falls within this interval. Note that this scheme yields 5\% false positives every time step. This criterion is examined independently at each echelon.

\subsubsection{Shared forecasting model}
Only echelon 1 updates the model by the same rule as in the  independent update scheme. If the forecasting model is updated, it is copied to all echelons. In this way, the forecasting model of echelon 1, which has the most accurate information about the demand, is shared to the other firms.

\subsection{Parameter settings and initial conditions}
We performed simulations for $N=5$ echelons, varying $L_{\rm train}$. The safety factor was set to $c=1.96$. As initial conditions, we set $\mu_0 = \mu_1=\cdots=\mu_N = 1.0$, $\sigma_0 = \sigma_1=\cdots=\sigma_N = 0.1$, and $I_1=\cdots=I_N = 1.196$.

\subsection{Constant policy}
To evaluate the efficacy of the retraining schemes, we prepared a baseline model, which is referred to as the constant policy. In this policy, each echelon's prediction is based on the Gaussian distribution, $\mathcal{N}(\bar{\mu},\sigma)$, which does not change over time. Here, $\bar{\mu}=1.25$ is a fixed parameter representing the long-term mean of demands ($\mu_0 \sim U(1/2,2)$), and we examined various values of $\sigma$ ranging from 0 to 1.2 in separate runs. This policy assumes an ideal situation in which each echelon knows the value of $\bar{\mu}$.

\section{Results}\label{sec:results}
\subsection{Emergence of bullwhip effect}
First, we present the average inventory level of each echelon in Fig. \ref{fig:inventory}. Upstream echelons had more inventory than downstream echelons except when the shared forecasting model algorithm was applied. 
A higher inventory level indicates that the echelon estimates larger variability in the demand signals (i.e., larger $\sigma_i$), 
resulting in ordering more to have more stock.

With the independent update algorithm, the inventory level became extremely high in upstream echelons when $L_{\rm train}$ was small (Fig. \ref{fig:inventory}(b)). In this case, unnecessarily frequent retraining perturbed the system too much, and its amplification caused the bullwhip effect. 

As expected, the shared forecasting model algorithm had the best performance in this respect. Because the forecasting model was shared to upstream echelons, they had the same estimate of demand variability, resulting in identical inventory levels.

\subsection{Lost sales opportunities}
We then examined the level of service achieved. In general, a lower inventory level leads to more stockouts and subsequent loss of sales opportunities.
Even if we consider the backlog of orders, stockouts causing additional costs are undesirable \cite{Croson2003,Disney2007}.
We quantified the number of lost sales opportunities of echelon $i$ at $t=T$ by $O_{i-1}(T) - S_{i}(T) (= \max{\{0, O_{i-1}(T) - I_{i}(T)\}})$. This value was divided by 1.25 (i.e., the long-term average of the demands) to compute the percentage of the lost sales opportunities (Fig. \ref{fig:l-s-o}).

The regular update scheme (Fig. \ref{fig:l-s-o}(a)) caused a substantial number of sales losses, which reached the maximum around $L_{\rm train}=T_{\rm{int}}/2$ where the inventory level was at its minimum (Fig. \ref{fig:inventory}(a)).
 
The independent update scheme (Fig. \ref{fig:l-s-o}(b)) suppressed the sales losses to a reasonable level while keeping the inventory level moderate when $L_{\rm{train}}$ was not very small (Fig. \ref{fig:inventory}(b)).

The shared forecasting model scheme significantly reduced the sales losses (Fig. \ref{fig:l-s-o}(c)) with a low inventory level (Fig. \ref{fig:inventory}(c)).

In Fig. \ref{fig:l-s-o}(a) and (b), echelon 2 or 3, not echelon 1, had the most sales losses. This is explained as follows. 
In these schemes, echelon 1 swiftly adjusted to the change in demand, and the upstream echelons followed after some time. 
When the demand increased, because echelons 2 and 3 did not receive an order larger than $I_1$, which is smaller than the inventory level of the other echelons, the increase in demand became difficult to detect, and adaptation to it was delayed.  

\subsection{Trade-off between inventory level and lost sales opportunities}
To further evaluate the performance of the three schemes (i.e., regular update, independent update, and shared forecasting model schemes), we consider a two-dimensional plane of the total inventory level (i.e., sum of $I_i,$ $i=1,...,N$) in the system, and the percentage of lost sales opportunities at echelon 1 (Fig. \ref{fig:pareto}). Note that the lost sales at echelon 1 are equal to those of the entire SC. Because these two indices have in a trade-off relationship, we must evaluate both at the same time to measure the performance. The lower left area in Fig. \ref{fig:pareto} corresponds to a lower inventory level and fewer lost sales opportunities (i.e., a set of better operating points). For various values of $L_{\rm train}$, the results of the three retraining schemes are plotted with the results of the constant  policy (red dashed line). Symbols located above this line indicate that the inventory control is inferior to that of the constant policy. 

The performance of the regular update scheme was comparable to that of the constant policy. 
In addition, the independent update and the shared forecasting model schemes provided substantially better operating points than the constant policy. In particular, the shared forecasting model scheme had the best performance. 
Moreover, the operating points of the shared forecasting model scheme were distributed over a small area, indicating the robustness of the scheme.

\section{Conclusion}\label{sec:conclusion}

We demonstrated that retraining a forecasting model can cause the bullwhip effect in an SC.
Furthermore, the shared forecasting model scheme effectively suppressed the bullwhip effect without increasing the loss of sales.
This scheme functioned robustly for various lengths of training data, which is an advantageous property for practical applications.

We illustrated the effect of model retraining for long-term demand fluctuations using a simple model. 
Because our findings are based on the intrinsic dynamic mechanisms of SCs, we believe that they generalize to other types of settings.

There are several factors that were ignored in the model, whose effects on our conclusions are discussed below. First, we ignored the lead time and its variability. 
If the lead time is variable, it should be reflected in the ordered amount (Eq. \eqref{eq:target}) \cite{Zhang2004,Kim2006,Disney2007}.
In this case, we suppose that the SC becomes more unstable and more vulnerable to perturbations by the retraining reported in this paper. 

Second, the effects of order batching \cite{Cachon2000,Potter2006} should be considered. 
In particular, when the echelons place orders asynchronously with different batch sizes, simple forecasting model sharing 
may not be sufficient to stabilize the SC. In this case, comprehensive information sharing, including the batching policy and current inventory level of downstream echelons, is required.

Finally, we did not include the temporal structure in the forecasting model because our focus was on retraining the models, which is primarily affected by the errors from the expectation. Thus, we did not consider seasonality \cite{Cho2012, Huang2017} and trends \cite{Chen2000}. However, if we regress these factors out from the original demand signal, the problem should remain essentially identical to that examined in this paper. 

In this paper, we proposed that sharing the forecasting model of  echelon 1 (i.e., a retailer) to upstream echelons benefits the entire SC.
In this scheme, only the retailer bears the cost of collecting information and performing retraining, although upstream suppliers benefit more  from the information. Thus, in practice, we should consider contracts among firms to incentivize retailers and downstream suppliers to share their models \cite{Lee2000,Cachon2001,Zhou2007,Choi2008,Viet2018}.

It should be noted that our results were based on a simple model, which is a general limitation of theoretical studies \cite{Syntetos2016}.
We believe that the effects of model retraining should be studied in greater detail with more practical settings in future research.
For instance, investigating the effects of model retraining on SC networks \cite{Tiwari2010, Lin2011,Baghalian2013} and
 more complex forecasting models for high-dimensional demand signals \cite{Ma2016} would be a useful future research direction.

\clearpage

\clearpage

\begin{figure*}[t]
\centering\includegraphics[width=160mm]{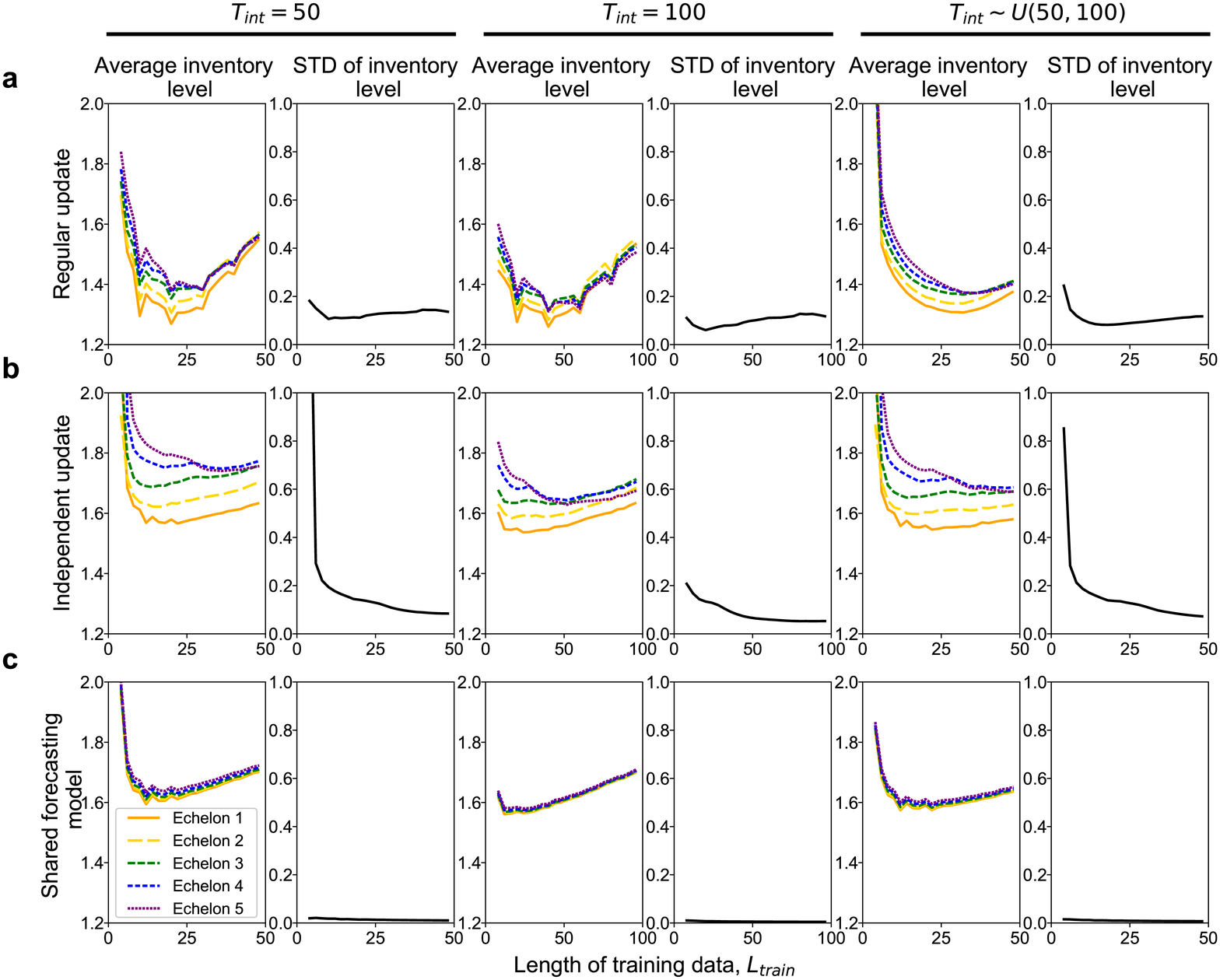}
\caption{Average inventory level at each echelon. 
(a) Regular update scheme. (b) Independent update scheme. (c) Shared forecasting model scheme.
In each panel, we varied the length of training data, $L_{\rm train}$.  
Simulations were performed for three types of intervals of demand change: $T_{\rm int} = 50$ (left),  $T_{\rm int} = 100$ (middle), and $T_{\rm int} \sim U(50, 100)$ (right). The standard deviation of the inventory level for 5 echelons was computed for each time step.
For each simulation condition, the results were averaged over $t=10^7$ steps. }
\label{fig:inventory}
\end{figure*}

\clearpage
\begin{figure*}[t]
\centering\includegraphics[width=160mm]{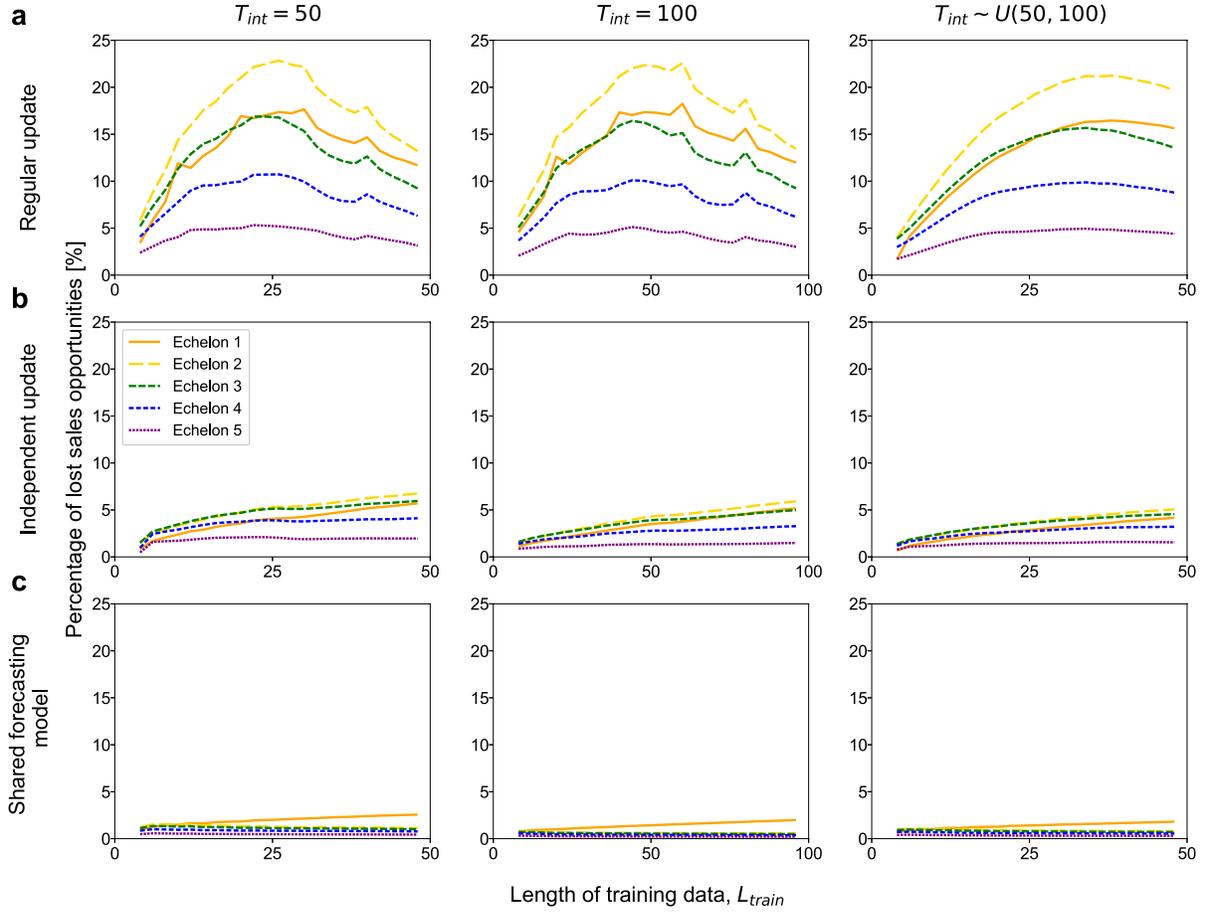}
\caption{Percentage of lost sales opportunities. (a) Regular update scheme. (b) Independent update scheme. (c) Shared forecasting model scheme.
In each panel, we varied the length of training data, $L_{\rm train}$. 
The simulations were performed for three types of intervals of demand change, i.e., $T_{\rm int} = 50$ (left),  $T_{\rm int} = 100$ (middle), and $T_{\rm int} \sim U(50, 100)$ (right).
For each simulation condition, the results were averaged over $t=10^7$ steps.  }
\label{fig:l-s-o}
\end{figure*}

\clearpage
\begin{figure*}[t]
\centering\includegraphics[width=160mm]{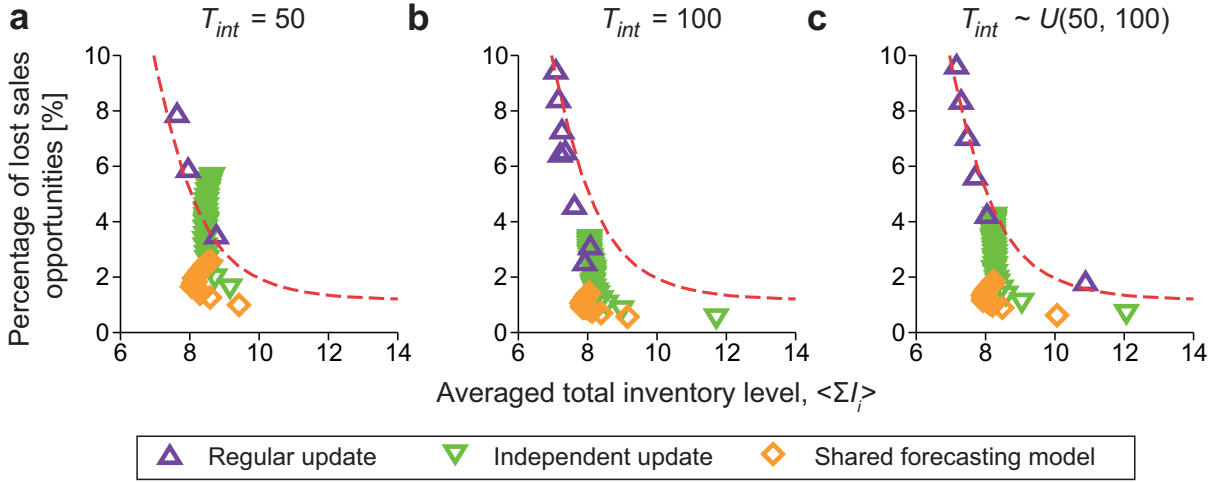}
\caption{Trade-off between the average inventory level (Fig. \ref{fig:inventory}) and the percentage of lost sales opportunities (Fig. \ref{fig:l-s-o}).
(a) $T_{\rm int} = 50$. (b) $T_{\rm int} = 100$. (c) $T_{\rm int} \sim U(50, 100)$.
Each symbol represents a simulation condition (i.e., a single $L_{\rm train}$ value) in Figs. \ref{fig:inventory} and \ref{fig:l-s-o}. The red dashed line represents the results of the constant policy.}
\label{fig:pareto}
\end{figure*}

\clearpage
\renewcommand{\figurename}{Supplementary Figure}
\renewcommand{\thefigure}{\Alph{section}\arabic{figure}}
\renewcommand{\thefigure}{\arabic{figure}}

\renewcommand{\theequation}{S\arabic{equation}}
\setcounter{section}{19}
\setcounter{equation}{0}
\setcounter{figure}{0}

\section*{Supporting Information}
\section*{Demands following a log-normal distribution}
In the main text, we assumed that the demands follow the Gaussian distribution. Here, to confirm that our conclusions are not sensitively influenced by the choice of the probability distribution, we performed simulations with the following log-normal distribution:
\begin{equation}
O_0 \sim \frac{1}{\sqrt{2\pi} \sigma_0 O_0}\exp{\left[ -\frac{(\ln{O_0} - \mu_0)}{2\sigma_0^2} \right]}.
\end{equation}
With this distribution, the logarithm of $O_0$ follows the Gaussian distribution, i.e., $\ln O_0 \sim \mathcal{N}(\mu_0, \sigma_0)$. 
The mean and variance of this distribution for fixed $\mu_0$ and $\sigma_0$, are expressed as 
$E[O_0 \mid \mu_0, \sigma_0] = \exp{[\mu_0 + \sigma_0^2/2]}$ and
$V[O_0 \mid \mu_0, \sigma_0] = \exp{[2\mu_0 + \sigma_0^2]}(\exp{[\sigma_0^2]}-1)$, respectively.

To implement the constant policy for this distribution, we numerically computed the expected value of $O_0$ (with varying $\mu_0(\sim U(0.5,2))$) to obtain $\bar{\mu}\approx 3.845$. This value was used to normalize the lost sales opportunities as well. The other settings of simulations were left unchanged from those used in the main text. 

The results are shown in Supporting Figs. 1--3. Because the mean and variance of the demands are larger than those used in the main text, the average inventory level and percentage of lost sales opportunities were larger than the results reported in Figs. 2--4 in the main text. However qualitative characteristics of the results were very similar, suggesting the robustness of our conclusions.

\clearpage

\begin{figure*}[t]
\centering\includegraphics[width=160mm]{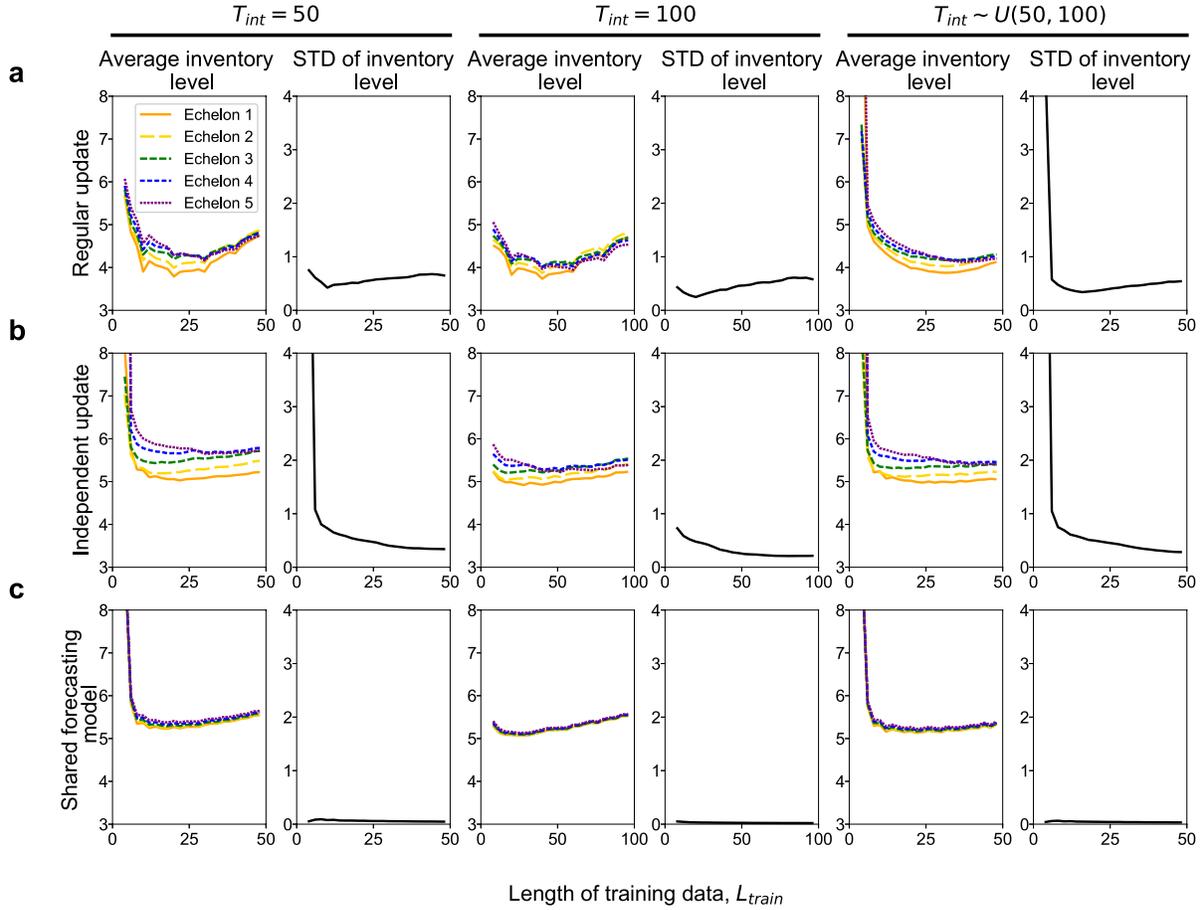}
\caption{Average inventory level at each echelon. 
(a) Regular update scheme. (b) Independent update scheme. (c) Shared forecasting model scheme.
In each panel, we varied the length of training data, $L_{\rm train}$. 
Simulations were performed for three types of intervals of demand change: $T_{\rm int} = 50$ (left),  $T_{\rm int} = 100$ (middle), and $T_{\rm int} \sim U(50, 100)$ (right). The standard deviation of the inventory level for 5 echelons was computed for each time step.
For each simulation condition, the results were averaged over $t=10^7$ steps. }
\label{fig:inventory}
\end{figure*}

\clearpage
\begin{figure*}[t]
\centering\includegraphics[width=160mm]{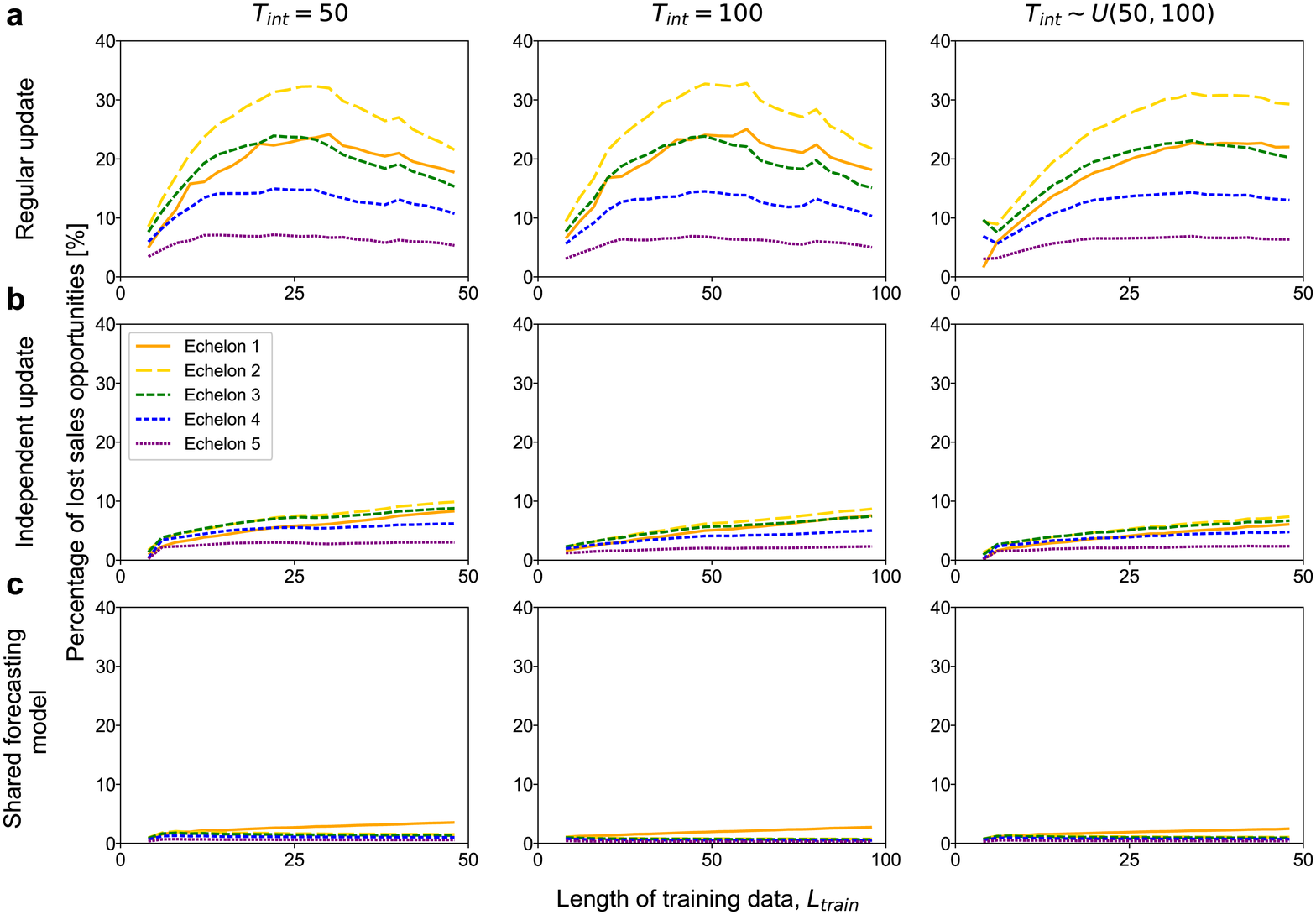}
\caption{Percentage of lost sales opportunities. (a) Regular update scheme. (b) Independent update scheme. (c) Shared forecasting model scheme.
In each panel, we varied the length of training data, $L_{\rm train}$. 
The simulations were performed for three types of intervals of demand change, i.e., $T_{\rm int} = 50$ (left),  $T_{\rm int} = 100$ (middle), and $T_{\rm int} \sim U(50, 100)$ (right).
For each simulation condition, the results were averaged over $t=10^7$ steps.  }
\label{fig:l-s-o}
\end{figure*}

\clearpage
\begin{figure*}[t]
\centering\includegraphics[width=160mm]{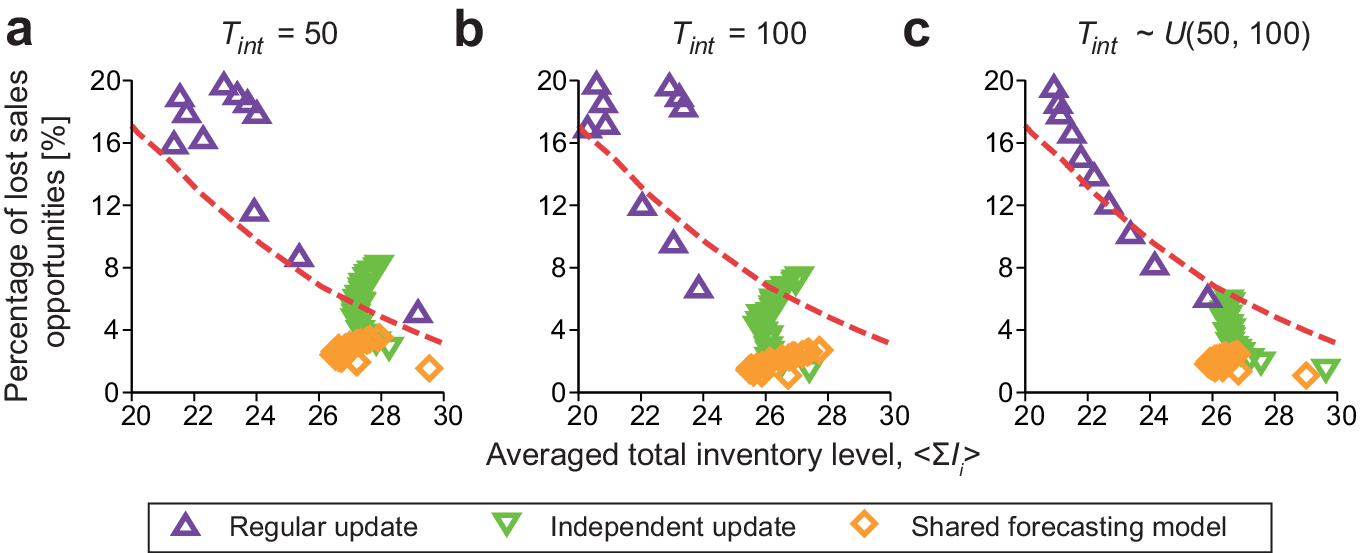}
\caption{Trade-off between the average inventory level (Supplementary Fig. \ref{fig:inventory}) and the percentage of lost sales opportunities (Supplementary Fig. \ref{fig:l-s-o}).
(a) $T_{\rm int} = 50$. (b) $T_{\rm int} = 100$. (c) $T_{\rm int} \sim U(50, 100)$.
Each symbol represents a simulation condition (i.e., a single $L_{\rm train}$ value) in Supplementary Figs. \ref{fig:inventory} and \ref{fig:l-s-o}. The red dashed line represents the results of the constant policy. }
\label{fig:pareto}
\end{figure*}

\end{document}